\documentclass[conference]{IEEEtran}
\IEEEoverridecommandlockouts
\usepackage{cite}
\usepackage{amsmath,amssymb,amsfonts}
\usepackage{algorithmic}
\usepackage{graphicx}
\usepackage{tabularx}
\usepackage{textcomp}
\usepackage{xcolor}
\usepackage{stfloats}
\usepackage{multirow}
\usepackage{subfigure}

\def\BibTeX{{\rm B\kern-.05em{\sc i\kern-.025em b}\kern-.08em
    T\kern-.1667em\lower.7ex\hbox{E}\kern-.125emX}}
\begin{document}

\title{A Novel 3D Non-stationary Localization-assisted ISAC Channel Model}

\author{Runruo Yang\textsuperscript{1,2}, Yang Wu\textsuperscript{1}, Jie Huang\textsuperscript{1,2*},  Cheng-Xiang Wang\textsuperscript{1,2*}
\\
\textsuperscript{1}National Mobile Communications Research Laboratory, School of Information Science and Engineering, 
\\Southeast University, Nanjing 210096, China.\\

\textsuperscript{2}Purple Mountain Laboratories, Nanjing 211111, China.\\
\textsuperscript{*}Corresponding Authors: Jie Huang and Cheng-Xiang Wang
\\
Email: \{yang\_rr, wu\_yang, j\_huang, chxwang\}@seu.edu.cn
}

\maketitle

\begin{abstract}
Integrated sensing and communication (ISAC) has attracted wide attention as an emerging application scenario for the sixth generation (6G) wireless communication system.
 In this paper, a novel three-dimensional (3D) non-stationary localization-assisted ISAC geometry-based stochastic model (GBSM) is proposed. The locations of the first-bounce scatterer and last-bounce scatterer in the communication channel can be estimated by the particle filter with the assistance of backscattering sensing. The important channel statistical properties of the proposed channel model are simulated and compared with the ray tracing (RT) results, including the delay spread, azimuth angle of departure/arrival (AAoD/AAoA) spread, and elevation angle of departure/arrival (EAoD/EAoA) spread. The simulation results of the proposed channel model show a good agreement with the RT results, which proves the correctness of the proposed channel model. Utilizing the localization parameters of scatterers, the proposed ISAC channel model can better map the real environment.
\end{abstract}


\section{Introduction}
With the development of wireless communication technology, the 6G wireless communication system requires both high data rate communication and high-precision sensing/localization to realize the mapping of the physical world and the digital world \cite{b1,rrb1}. For the reason that sensing and communication have commonalities in terms of utilized frequency bands, waveform design, and hardware platform \cite{b2,rb1,rb2}, ISAC has received extensive attention as an emerging 6G application scenario. The channel model plays an essential role in system design, resource allocation, and performance optimization \cite{b3}. Therefore, in order to realize the mutual assistance between sensing and communication of ISAC system, it is necessary to study the ISAC channel.

The localization method realized on the channel model, and it has been studied in the literatures. 
In \cite{b5}, a millimeter wave (mmWave) beam domain communication channel model was proposed, and the maximum-likelihood (ML) classifier was utilized for receiver (Rx) localization. 
In \cite{b6}, a multipath components (MPC) tracking algorithm based on the GBSM was proposed, and made a comparison between the simulation results and the measurement data.
In \cite{b7}, a method for tracking user motion using multipath phase information by extended Kalman filter was proposed. Compared with the measured data in line-of-sight (LoS) scenario, this method was proved to be able to track the user motion trajectory well.
However, scatterers in the environment cannot be located by these methods.
In \cite{b8}, a map-free localization method for indoor scenario was proposed, and the estimated transmitter (Tx) and clusters locations were compared with measurement results. 
However, this method can only be used to locate Tx and scatterers in the LoS path or one-bounce path, and it does not apply to scenarios with multi-bounce paths.
In \cite{b9}, an indoor environment mapping method based on sensing backscattering channel was proposed. 
However, the LoS path was required between the target and transceiver, and scatterers can not be located. 
Localization methods mentioned above can only locate the user or single bounce clusters. 
For multi-bounce clusters within MPCs, the location of the scatterer cannot be estimated by the above methods.

Channel models have also been studied in many literatures. 
In \cite{b3}, a 6G pervasive channel model (6GPCM) was proposed, in which positions of scatterers were randomly generated by ellipsoid Gaussian distribution. 
A 3D non-stationary unmanned aerial vehicle (UAV) multiple-input multiple-output (MIMO) channel model was proposed in \cite{b11}. Angle parameters were modelled by von Mises-Fisher distribution.
In \cite{b12}, a mobile-to-mobile two-ring non-stationary channel model was proposed, which assumed that the scatterers were regularly distributed on the Tx and Rx rings.
In \cite{c1,c2}, the 3D non-stationary UAV MIMO GBSMs were proposed, in which scatterers were distributed on the cylinder surface at Rx side.
A high-speed train channel model was proposed in \cite{c3}, and the scatterers were located on the ellipse with Tx and Rx as the foci.
However, in the above channel models, the locations of scatterers are characterized by specific random distributions or the scatterers are distributed on the surface of a specific geometry.
The spatial positions of the scatterers generated by the above methods can not reflect the real positions of the objects in the physical environment.

In this paper, a novel 3D non-stationary localization-assisted ISAC channel model is proposed. Through the combination of communication and sensing, the locations of multi-bounce scatterers in the channel can be determined with the help of the particle filter. Considering the non-stationarity of the channel model, the locations of scatterers will change with time, and the resampling process of the particle filter can reflect this property. The process of resampling also implies the birth-death of scatterers in the channel model. 

The remainder of this paper is organized as follows. 
In Section~\uppercase\expandafter{\romannumeral2}, a novel 3D non-stationary localization-assisted ISAC channel model is proposed. In Section~\uppercase\expandafter{\romannumeral3}, the channel statistical properties are analyzed, including the delay spread, AAoD/AAoA spread, and EAoD/EAoA spread. Section~\uppercase\expandafter{\romannumeral4} presents the results and analysis. Finally, the paper is concluded in Section~\uppercase\expandafter{\romannumeral5}.

\section{A Novel 3D Localization-assisted ISAC GBSM}
\subsection{Channel Model Description}
\begin{figure}[tb]
	\centering{\includegraphics[width=0.45\textwidth]{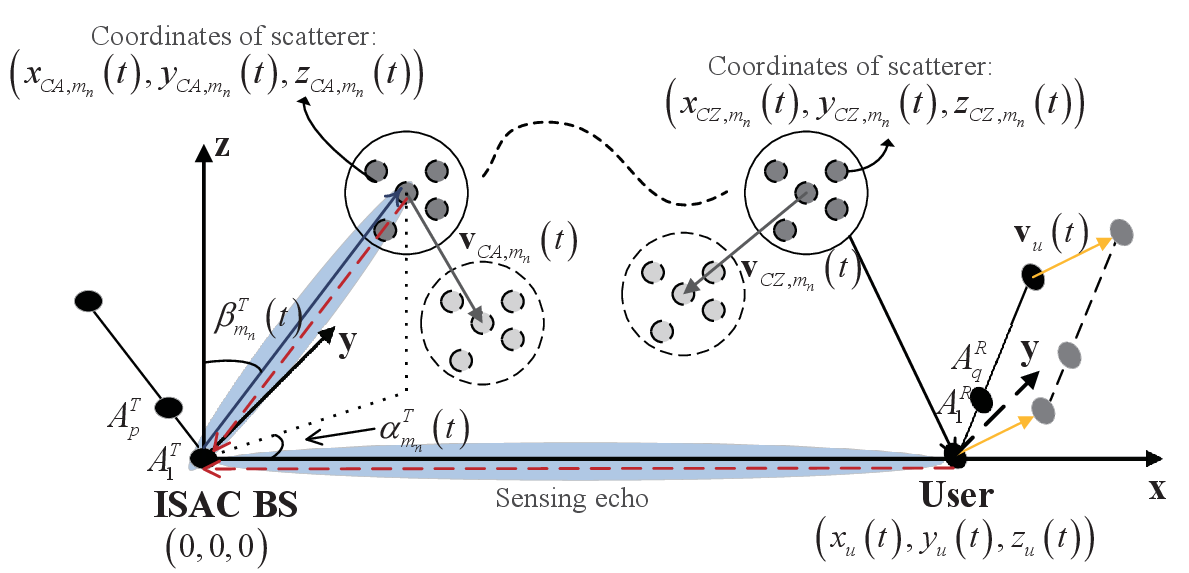}}
	\caption{ISAC channel model.}
	\label{CM}
\end{figure}

The ISAC channel model is composed of a mono-static sensing channel and a communication channel. 
There is an ISAC base station (BS) at Tx, and the Rx is a communication user. 
A transmit antenna array and a receive antenna array are both equipped at the ISAC BS.
The ISAC BS can transmit sensing and communication signals, and receive the sensing echo signal. The Rx possesses an antenna array for receiving communication signals. 
Since the communication transmitting and receiving arrays are located at different positions, the communication channel can also be seen as a bi-static channel.
Locations of the first-bounce (last-bounce) scatterers within the channel model can be determined by delay, AAoD (AAoA), and EAoD (EAoA) parameters, which are indispensable parameters of the channel model.
Fig. \ref{CM} depicts the proposed ISAC channel model.
\subsubsection{	Mono-static Sensing Channel Model}
The ISAC BS transmits pilot symbols or pseudo-noise (PN) sequences to sense the obstacles in the physical environment distributed around the ISAC BS. These obstacles will be abstracted as scatterers in the sensing channel model. 
The scatterer will be sensed by the ISAC BS, if there is a LoS path between the ISAC BS and the scatterer. 
The distance, angles, and velocity of the scatterer can be obtained by the sensing channel. 
The mono-static sensing echo channel impulse response (CIR) is given by
\begin{equation}\label{sensing}
\begin{aligned}
{h^{mo}}\left( {t,\tau } \right) =& \sum\limits_{i = 1}^{I\left( t \right)} {\sum\limits_{j = 1}^{{J_i}} {\sqrt {{G_{{j_i}}}\left( t \right)} } } {e^{j2\pi {f_c}{\tau _{{j_i}}}\left( t \right)}}{e^{j2\pi {f_{D,{j_i}}}\left( t \right){\tau _{{j_i}}}\left( t \right)}}
\\ & A\left( {{\alpha _{{j_i}}}\left( t \right),{\beta _{{j_i}}}\left( t \right)} \right)\delta \left( {\tau  - {\tau _{{j_i}}}\left( t \right)} \right)
\end{aligned} 
\end{equation}
where ${{G}_{{{j}_{i}}}}\left( t \right)={{\lambda }^{2}}{{{\varepsilon }_{RCS,{{j}_{i}}}}}/{ 64{{\pi }^{3}}{{d}_{{{j}_{i}}}^{4}}{{\left( t \right)}} }$ is the complex channel gain, ${{\varepsilon }_{RCS,{{j}_{i}}}}$ is the radar cross section (RCS) of the $j$th scatterer in the $i$th cluster, $\lambda$ is the wavelength, ${{d}_{{{j}_{i}}}}\left( t \right)$ is the distance between ISAC BS and the $j$th scatterer, ${{f}_{c}}$ represents the carrier frequency, ${{\tau }_{{{j}_{i}}}}\left( t \right)$ represents the delay, and ${{f}_{D,{{j}_{i}}}}\left( t \right)$ is Doppler shift of the $j$th scatterer. 
The steering vector of ISAC BS is represented as $A\left( {{\alpha _{{j_i}}}\left( t \right),{\beta _{{j_i}}}\left( t \right)} \right)$, where ${\alpha_{{j_i}}}\left( t \right)$ and ${\beta_{{j_i}}}\left( t \right) $ are the azimuth and elevation angles of the $j$th scatterer, respectively. 
The delay of the echo path is expressed as ${{\tau }_{{{j}_{i}}}}\left( t \right)=2{{d}_{{{j}_{i}}}}\left( t \right)/c$, and the Doppler shift is ${{f}_{D,{{j}_{i}}}}\left( t \right)=2{{v}_{{{j}_{i}}}}\left( t \right)/\lambda \ $, where $c$ and ${{v}_{{{j}_{i}}}}\left( t \right)$ are the velocities of light and the $j$th scatterer, respectively.  

\subsubsection{	Communication Channel Model}
The communication channel refers to the channel between ISAC BS and user. 
At the same time, this channel can also be regarded as a bi-static channel, in which ISAC BS and users are located at different locations. 
The channel model consists of LoS $h_{qp}^{LoS}\left( t,\tau  \right)$ and non-LoS (NLoS) $h_{qp,{{m}_{n}}}^{NLoS}\left( t,\tau  \right)$ components.
The communication CIR can be represented as
\begin{equation}\label{com}
\begin{aligned}
h_{qp}^{c}\left( {t,\tau } \right) =& \sqrt {\frac{{K\left( t \right)}}{{K\left( t \right) + 1}}} h_{qp}^{LoS}\left( {t,\tau } \right)+\\& \sqrt {\frac{1}{{K\left( t \right) + 1}}} \sum\limits_{n = 1}^{{N_c}\left( t \right)} {\sum\limits_{m = 1}^{{M_n}\left( t \right)} {h_{qp,{m_n}}^{NLoS}\left( {t,\tau } \right)} } 
\end{aligned} 
\end{equation}
where $K\left( t \right)$ denotes the Rician factor, ${{N}_{c}}\left( t \right)$ denotes the number of clusters at time $t$, and ${{M}_{n}}\left( t \right)$ denotes the number of scatterers in the $n$th cluster.

The LoS component $h_{qp}^{LoS}\left( t,\tau  \right)$ can be calculated as
\begin{small}
\begin{equation}\label{com}
\begin{aligned}
&h_{qp}^{LoS}\left( {t,\tau } \right) = {\left[ \begin{array}{l}
	{F_{q,V}}\left( {\beta _L^R\left( t\right),\alpha _L^R\left( t\right)} \right)\\
	{F_{q,H}}\left( {\beta _L^R\left( t\right),\alpha _L^R\left( t\right)} \right)
	\end{array} \right]^T}\left[ \begin{array}{ccc}
{e^{j\xi _L^{VV}}} & 0\\
0 &  - {e^{j\xi _L^{HH}}}
\end{array} \right] \\&
\left[ \begin{array}{l}
{F_{p,V}}\left( {\beta _L^T\left( t\right),\alpha _L^T\left( t\right)} \right)\\
{F_{p,H}}\left( {\beta _L^T\left( t\right),\alpha _L^T\left( t\right)} \right)
\end{array} \right]{e^{j2\pi {f_c}\tau _{qp}^{LoS}\left( t \right)}}\delta \left( {\tau  - \tau _{qp}^{LoS}\left( t \right)} \right)
\end{aligned} 
\end{equation}
\end{small}
where ${\left\{  \cdot  \right\}^T}$ means transposition, $\delta \left( \cdot  \right)$ means Dirac delta function, ${F_{p\left( q \right),V}}$ and ${F_{p\left( q \right),H}}$ are vertical and horizontal polarizations of the antenna patterns for ISAC BS (Rx), respectively. 
The variables $\alpha _L^T\left( t\right)$ ($\alpha _L^R\left( t\right)$) and $\beta _L^T\left( t\right)$ ($\beta _L^R\left( t\right)$) are the AAoD (AAoA) and EAoD (EAoA) of LoS path, respectively.
The initial phases $\xi _L^{HH}$ and $\xi _L^{VV}$ follow the uniform distribution of $[0, 2\pi]$.
The variable $\tau _{qp}^{LoS}\left( t \right)$ represents the delay of LoS path between the $p$th transmit antenna and the $q$th receive~antenna.

The NLoS component is defined as  
	\begin{equation}
	\begin{aligned}
	h_{qp,{m_n}}^{NLoS}\left( {t,\tau } \right) = &{\left[ \begin{array}{l}
		{F_{q,V}}\left( {\beta _{{m_n}}^R\left( t\right),\alpha _{{m_n}}^R\left( t\right)} \right)\\
		{F_{q,H}}\left( {\beta _{{m_n}}^R\left( t\right),\alpha _{{m_n}}^R\left( t\right)} \right)
		\end{array} \right]^T}\\&
	\left[ 
	\begin{array}{cc}
	{e^{j\xi _{{m_n}}^{VV}}} & \sqrt {\mu \kappa _{{m_n}}^{ - 1}\left( t \right)} {e^{j\xi _{{m_n}}^{VH}}}\\
	\sqrt {\kappa _{{m_n}}^{ - 1}\left( t \right)} {e^{j\xi _{{m_n}}^{HV}}} & \sqrt \mu  {e^{j\xi _{{m_n}}^{HH}}}
	\end{array} \right]\\&
	\left[ \begin{array}{l}
	{F_{p,V}}\left( {\beta _{{m_n}}^T\left( t\right),\alpha _{{m_n}}^T\left( t\right)} \right)\\
	{F_{p,H}}\left( {\beta _{{m_n}}^T\left( t\right),\alpha _{{m_n}}^T\left( t\right)} \right)
	\end{array} \right]\\&
	\sqrt {{P_{qp,{m_n}}}\left( t \right)} {e^{j2\pi {f_c}\tau _{qp,{m_n}}^{NLoS}\left( t \right)}}\delta \left( {\tau  - \tau _{qp,{m_n}}^{NLoS}\left( t \right)} \right)
	\label{cir_nLoS}
	\end{aligned}
	\end{equation}
where, $\mu$ is the co-polarization imbalance, ${\kappa _{{m_n}}}$ is the cross-polarization power ratio, $\alpha _{{m_n}}^{T\left( R \right)}\left( t \right)$ and $\beta _{{m_n}}^{T\left( R \right)}\left( t \right)$ are AAoD (AAoA) and EAoD (EAoA) of the $m$th ray in the $n$th cluster at time $t$,
 ${{P_{qp,{m_n}}}\left( t \right)}$ is the power of path, and $\tau _{qp,{{m}_{n}}}^{NLoS}\left( t \right)$ is the delay of NLoS path between the $p$th transmit antenna and the $q$th receive antenna via the $m$th scatterer of the $n$th cluster. The delay is given by $\tau _{qp,{m_n}}^{NLoS}\left( t \right) = {{{D_{qp,{m_n}}}\left( t \right)}\mathord{\left/{\vphantom {{{D_{qp,{m_n}}}\left( t \right)} c}} \right.\kern-\nulldelimiterspace} c} + \tilde \tau \left( t \right)$, where ${{D_{qp,{m_n}}}\left( t \right)}$ is the distance of NLoS path between the $p$th transmit antenna and the $q$th receive antenna, and $\tilde \tau \left( t \right)$ is the virtual delay. The distance ${{D_{qp,{m_n}}}\left( t \right)}$ can be computed according to 
\begin{equation}
\begin{aligned}
{D_{qp,{m_n}}}\left( t \right) = \left| {{{\bf{D}}_{p,{m_n}}}\left( t \right)} \right| + \left| {{{\bf{D}}_{q,{m_n}}}\left( t \right)} \right|
\end{aligned} 
\end{equation}
where ${{{\bf{D}}_{p,{m_n}}}\left( t \right)}$ and ${{{\bf{D}}_{q,{m_n}}}\left( t \right)}$ are the distance vectors from $p$th transmit antenna to the $m$th scatterer and from the $q$th receive antenna to the $m$th scatterer, respectively. These distance vectors are given as
\begin{equation}
\begin{aligned}
{{\bf{D}}_{p,{m_n}}}\left( t \right) = {\bf{D}}_{{m_n}}^T\left( t \right) - {\bf{A}}_p^T
\end{aligned} 
\end{equation}
\begin{equation}
\begin{aligned}
{{\bf{D}}_{q,{m_n}}}\left( t \right) = {\bf{D}}_{{m_n}}^R\left( t \right) - {\bf{A}}_q^R
\end{aligned} 
\end{equation}
where ${\bf{D}}_{{m_n}}^{T(R)}\left( t \right)$ is the distance vector from the first antenna at the Tx (Rx) to the $m$th scatterer at time $t$, and ${\bf{A}}_p^T ({\bf{A}}_q^R)$ is the vector from the first antenna of the transmitter (receiver) to the $p$th ($q$th) antenna. The vectors ${\bf{D}}_{{m_n}}^{T}\left( t \right)$ and ${\bf{D}}_{{m_n}}^R\left( t \right)$ can be calculated as
\begin{equation}
\begin{aligned}
{\bf{D}}_{{m_n}}^T\left( t \right) = D_{{m_n}}^T\left( t \right){\left[ \begin{array}{c}
	\cos \left( {\beta _{{m_n}}^T\left( t \right)} \right)\cos \left( {\alpha _{{m_n}}^T\left( t \right)} \right)\\
	\cos \left( {\beta _{{m_n}}^T\left( t \right)} \right)\sin \left( {\alpha _{{m_n}}^T\left( t \right)} \right)\\
	\sin \left( {\beta _{{m_n}}^T\left( t \right)} \right)
	\end{array} \right]}
\end{aligned} 
\end{equation}
\begin{equation}
\begin{aligned}
{\bf{D}}_{{m_n}}^R\left( t \right) = D_{{m_n}}^R\left( t \right)\left[ \begin{array}{c}
\cos \left( {\beta _{{m_n}}^R\left( t \right)} \right)\cos \left( {\alpha _{{m_n}}^R\left( t \right)} \right)\\
\cos \left( {\beta _{{m_n}}^R\left( t \right)} \right)\sin \left( {\alpha _{{m_n}}^R\left( t \right)} \right)\\
\sin \left( {\beta _{{m_n}}^R\left( t \right)} \right)
\end{array} \right]
\end{aligned} 
\end{equation}
where ${{D}}_{{m_n}}^{T}\left( t \right)$ and ${{D}}_{{m_n}}^{R}\left( t \right)$ are the amplitudes of the distance vectors.
The locations of multi-bounce scatterers in the channel can be estimated by particle filter.  
The delay and angle parameters of scatterers mentioned above can also be obtained, rather than by random generation.

\subsection{Particle Filter-based Localization}
The first-bounce scatterers of communication channel model, which is distributed around the Tx, can be sensed by ISAC BS.
With the assistance of sensing, the last-bounce scatterers in the communication channel can be located and tracked by particle filter. The state vector of particle filter at time ${{t}_{k}}$ is given as 
\begin{equation}
\begin{aligned}
\mathbf{s}\left( {{t}_{k}} \right)={{\left[ {{\mathbf{s}}_{u}}\left( {{t}_{k}} \right),{{\mathbf{s}}_{CA}}\left( {{t}_{k}} \right),{{\mathbf{s}}_{CZ}}\left( {{t}_{k}} \right) \right]}^{T}}
\end{aligned} 
\end{equation}
 where ${{\mathbf{s}}_{u}}\left( {{t}_{k}} \right)$ is the state vector of user, ${{\bf{s}}_{CA}}\left( {{t_k}} \right) = {\left[ {...,{{\bf{s}}_{CA,{m_n}}}\left( {{t_k}} \right),...} \right]^T}  $ and ${{\bf{s}}_{CZ}}\left( {{t_k}} \right) = {\left[ {...,{{\bf{s}}_{CZ,{m_n}}}\left( {{t_k}} \right),...} \right]^T}$, ($ {m = 1,\ldots,{M_n}\left( {{t_k}} \right),n = 1,\ldots,{N_c}\left( {{t_k}} \right)}$) are the state vectors of the first-bounce scatterer and the last-bounce scatterer, respectively. 
 The vectors ${{\mathbf{s}}_{u}}\left( {{t}_{k}} \right)$, ${{\mathbf{s}}_{CA,{{m}_{n}}}}\left( {{t}_{k}} \right)$, and ${{\mathbf{s}}_{CZ,{{m}_{n}}}}\left( {{t}_{k}} \right)$ can be represented as 
 \begin{subequations}
 	\begin{align}
 	{{\mathbf{s}}_{u}}\left( {{t}_{k}} \right)=&{{\left[ {{\mathbf{p}}_{u}}\left( {{t}_{k}} \right),{{\mathbf{v}}_{u}}\left( {{t}_{k}} \right) \right]}^{T}} \\
 	{{\mathbf{s}}_{CA,{{m}_{n}}}}\left( {{t}_{k}} \right)=&{{\left[ {{\mathbf{p}}_{CA,{{m}_{n}}}}\left( {{t}_{k}} \right),{{\mathbf{v}}_{CA,{{m}_{n}}}}\left( {{t}_{k}} \right) \right]}^{T}}  \\
 {{\mathbf{s}}_{CZ,{{m}_{n}}}}\left( {{t}_{k}} \right)=&{{\left[ {{\mathbf{p}}_{CZ,{{m}_{n}}}}\left( {{t}_{k}} \right),{{\mathbf{v}}_{CZ,{{m}_{n}}}}\left( {{t}_{k}} \right) \right]}^{T}}.
 	\end{align}
 \end{subequations}
The location vectors of user, the first-bounce scatterer, and the last-bounce scatterer are represented as
 \begin{subequations}
	\begin{align}
	{{\mathbf{p}}_{u}}\left( {{t}_{k}} \right)=&{{\left[ {{x}_{u}}\left( {{t}_{k}} \right),{{y}_{u}}\left( {{t}_{k}} \right),{{z}_{u}}\left( {{t}_{k}} \right) \right]}^{T}} \\
	{{\mathbf{p}}_{CA,{{m}_{n}}}}\left( {{t}_{k}} \right)=&{{\left[ {{x}_{CA,{{m}_{n}}}}\left( {{t}_{k}} \right),{{y}_{CA,{{m}_{n}}}}\left( {{t}_{k}} \right),{{z}_{CA,{{m}_{n}}}}\left( {{t}_{k}} \right) \right]}^{T}}  \\
	{{\mathbf{p}}_{CZ,{{m}_{n}}}}\left( {{t}_{k}} \right)=&{{\left[ {{x}_{CZ,{{m}_{n}}}}\left( {{t}_{k}} \right),{{y}_{CZ,{{m}_{n}}}}\left( {{t}_{k}} \right),{{z}_{CZ,{{m}_{n}}}}\left( {{t}_{k}} \right) \right]}^{T}}
	\end{align}
\end{subequations}
where ${{x}_{u}}\left( {{t}_{k}} \right)$, ${{y}_{u}}\left( {{t}_{k}} \right)$, and ${{z}_{u}}\left( {{t}_{k}} \right)$ are the coordinates of user in rectangular coordinate system, respectively. Similarly, ${{x}_{CA\left( CZ \right),{{m}_{n}}}}\left( {{t}_{k}} \right)$, ${{y}_{CA\left( CZ \right),{{m}_{n}}}}\left( {{t}_{k}} \right)$, and ${{z}_{CA\left( CZ \right),{{m}_{n}}}}\left( {{t}_{k}} \right)$ are the coordinates of the first-bounce (last-bounce) scatterer, respectively. 
The velocity vectors of user, the first-bounce scatterer, and the last-bounce scatterer are represented as
 \begin{subequations}
	\begin{align}
	{{\mathbf{v}}_{u}}\left( {{t}_{k}} \right)=&{{\left[ v_{u}^{x}\left( {{t}_{k}} \right),v_{u}^{y}\left( {{t}_{k}} \right),v_{u}^{z}\left( {{t}_{k}} \right) \right]}^{T}} \\
	{{\mathbf{v}}_{CA,{{m}_{n}}}}\left( {{t}_{k}} \right)=&{{\left[ v_{CA,{{m}_{n}}}^{x}\left( {{t}_{k}} \right),v_{CA,{{m}_{n}}}^{y}\left( {{t}_{k}} \right),v_{CA,{{m}_{n}}}^{z}\left( {{t}_{k}} \right) \right]}^{T}}  \\
	{{\mathbf{v}}_{CZ,{{m}_{n}}}}\left( {{t}_{k}} \right)=&{{\left[ v_{CZ,{{m}_{n}}}^{x}\left( {{t}_{k}} \right),v_{CZ,{{m}_{n}}}^{y}\left( {{t}_{k}} \right),v_{CZ,{{m}_{n}}}^{z}\left( {{t}_{k}} \right) \right]}^{T}}
	\end{align}
\end{subequations}
where $v_{u}^{x}\left( {{t}_{k}} \right)$, $v_{u}^{y}\left( {{t}_{k}} \right)$, and $v_{u}^{z}\left( {{t}_{k}} \right)$ are the magnitudes of the user velocity vector along the $x$, $y$, and $z$ axis, respectively. Symbols $v_{CA\left( CZ \right),{{m}_{n}}}^{x}\left( {{t}_{k}} \right)$, $v_{CA\left( CZ \right),{{m}_{n}}}^{y}\left( {{t}_{k}} \right)$, and $v_{CA\left( CZ \right),{{m}_{n}}}^{z}\left( {{t}_{k}} \right)$ are the speeds of the first-bounce (last-bounce) scatterer along the $x$, $y$, and $z$ axis, respectively.
Utilizing the state values, the distance parameters of channel model can be calculated as
 \begin{equation}
 \begin{aligned}
\hat D_{{m_n}}^T\left( {{t_k}} \right) = \left[{{ {{x^2_{CA,{m_n}}}\left( {{t_k}} \right)} }} + {{ {{y^2_{CA,{m_n}}}\left( {{t_k}} \right)} }} +
{{ {{z^2_{CA,{m_n}}}\left( {{t_k}}  \right)}}}\right]^{\frac{1}{2}} 
 \end{aligned} 
 \end{equation}
 \begin{equation}
\begin{aligned}
\hat D_{{m_n}}^R\left( {{t_k}} \right) =& \left[{{\left( {{x_{CZ,{m_n}}}\left( {{t_k}} \right) - {x_u}\left( {{t_k}} \right)} \right)}^2}+{{\left( {{y_{CZ,{m_n}}}\left( {{t_k}} \right)-  }\right.}}\right.\\&\left.{{\left.{
 {y_u}\left( {{t_k}} \right)} \right)}^2}  +{{\left( {{z_{CZ,{m_n}}}\left( {{t_k}} \right) - {z_u}\left( {{t_k}} \right)} \right)}^2}\right]^{\frac{1}{2}}  
\end{aligned} 
\end{equation}
where $\left({\hat  \cdot }\right)$ means the estimated value.
We assume ${\Upsilon _{1}}=\arctan \left( {\frac{{{y_{CA,{m_n}}}\left( {{t_k}} \right)}}{{{x_{CA,{m_n}}}\left( {{t_k}} \right)}}} \right)$, the estimated AAoD is calculated as 
\begin{equation}
\begin{aligned}
\begin{split}
\hat \alpha _{{m_n}}^T\left( {{t_k}} \right)= \left \{
\begin{array}{ll}
{\Upsilon _{1}},                    & {{x_{CA,{m_n}}}\left( {{t_k}} \right) > 0,{y_{CA,{m_n}}}\left( {{t_k}} \right) \mathbin{\lower.3ex\hbox{$\buildrel>\over
			{\smash{\scriptstyle<}\vphantom{_x}}$}} 0}\\
{\Upsilon _{1}}+\pi,   & {{x_{CA,{m_n}}}\left( {{t_k}} \right) < 0,{y_{CA,{m_n}}}\left( {{t_k}} \right) > 0}\\
{\Upsilon _{1}}-\pi,                              & {{x_{CA,{m_n}}}\left( {{t_k}} \right) < 0,{y_{CA,{m_n}}}\left( {{t_k}} \right) < 0}.
\end{array}
\right.
\end{split}
\end{aligned}
\end{equation}
Similarly, we assume the ${\Upsilon _{2}} = \arctan \left( {\frac{{{y_{CZ,{m_n}}}\left( {{t_k}} \right) - {y_u}\left( {{t_k}} \right)}}{{{x_{CZ,{m_n}}}\left( {{t_k}} \right) - {x_u}\left( {{t_k}} \right)}}} \right)$,
the estimated AAoA can be calculated as 
\begin{equation}
\begin{aligned}
\begin{split}
\hat \alpha _{{m_n}}^R\left( {{t_k}} \right)= \left \{
\begin{array}{ll}
{\Upsilon _{2}},                    & \begin{array}{l}
{x_{CZ,{m_n}}}\left( {{t_k}} \right) - {x_u}\left( {{t_k}} \right) > 0,\\
{y_{CZ,{m_n}}}\left( {{t_k}} \right) - {y_u}\left( {{t_k}} \right) \mathbin{\lower.3ex\hbox{$\buildrel>\over
		{\smash{\scriptstyle<}\vphantom{_x}}$}} 0
\end{array}\\
{\Upsilon _{2}}+\pi,   & \begin{array}{l}
{x_{CZ,{m_n}}}\left( {{t_k}} \right) - {x_u}\left( {{t_k}} \right) < 0,\\
{y_{CZ,{m_n}}}\left( {{t_k}} \right) - {y_u}\left( {{t_k}} \right) > 0
\end{array}\\
{\Upsilon _{2}}-\pi,                              & \begin{array}{l}
{x_{CZ,{m_n}}}\left( {{t_k}} \right) - {x_u}\left( {{t_k}} \right) < 0,\\
{y_{CZ,{m_n}}}\left( {{t_k}} \right) - {y_u}\left( {{t_k}} \right) < 0.
\end{array}
\end{array}
\right.
\end{split}
\end{aligned}
\end{equation}
The EAoD and EAoA can be estimated as 
 \begin{equation}
\begin{aligned}
\hat \beta _{{m_n}}^T\left( {{t_k}} \right) = \arccos \left( {\frac{{{z_{CA,{m_n}}}\left( {{t_k}} \right)}}{{\hat D_{{m_n}}^T\left( {{t_k}} \right)}}} \right) 
\end{aligned} 
\end{equation}

 \begin{equation}
\begin{aligned}
\hat \beta _{{m_n}}^R\left( {{t_k}} \right) = \arccos \left( {\frac{{{z_{CZ,{m_n}}}\left( {{t_k}} \right) - {z_u}\left( {{t_k}} \right)}}{{\hat D_{{m_n}}^R\left( {{t_k}} \right)}}} \right).
\end{aligned} 
\end{equation}

Utilizing the particle filter, the state vector is recursive in the time domain, and the location and velocity vectors of user and scatterers at different times can be obtained. For the reason that the mono-static channel model and communication channel model has an intersection, the initial state vector can be determined jointly by both of them. It is assumed that the locations of Tx and Rx, and the velocity of Rx at initial time are known. To acquire the communication channel information, the ISAC BS transmits the pilot signal to the Rx at the initial time. At the same time, the receiving array of the ISAC BS will receive the echo signals backscattered from the scatterers around the ISAC BS. The received signals of Rx and receiving array of the ISAC BS are processed separately to obtain the communication channel information and mono-static sensing channel information. The initial state of the first-bounce scatterer ${{\mathbf{s}}_{CA}}\left( {{t}_{1}} \right)$ can be determined by the intersection of mono-static sensing channel and communication channel. The state vector of the last-bounce scatterer at initial time ${{\mathbf{s}}_{CZ}}\left( {{t}_{1}} \right)$ is determined by the communication channel individually. 
After the state vector is initialized, it needs to recurse in the time domain. 
The state transition equation from time $t_{k-1}$ to $t_{k}$ can be represented as
 \begin{equation}
\begin{aligned}
\mathbf{s}\left( {{t}_{k}} \right)=\mathbf{As}\left( {{t}_{k-1}} \right)+{{\mathbf{n}}_{s}}\left( {{t}_{k}} \right)
\end{aligned} 
\end{equation}
where $\mathbf{A}$ is state transition matrix, ${{\mathbf{n}}_{s}}\left( {{t}_{k}} \right)$ is system noise, which follows the Gaussian random distribution with zero mean value. The state transition matrix $\mathbf{A}$ can be computed as
 \begin{equation}
\begin{aligned}
{\bf{A}} = \left[ {\begin{array}{*{20}{c}}
	{\bf{Q}}&{\bf{0}}&{...}&{\bf{0}}&{\bf{0}}\\
	{\bf{0}}&{\bf{Q}}&{...}&{\bf{0}}&{\bf{0}}\\
	{...}&{...}&{...}&{...}&{...}\\
	{\bf{0}}&{\bf{0}}&{...}&{\bf{Q}}&{\bf{0}}\\
	{\bf{0}}&{\bf{0}}&{...}&{\bf{0}}&{\bf{Q}}
	\end{array}} \right]
\end{aligned} 
\end{equation}
where 
 \begin{equation}
\begin{aligned}
{\bf{Q}} = \left[ \begin{array}{cccccc}
1& 0 &0& {T_{\rm{s}}}& 0& 0\\
0 &1& 0& 0& {T_s}& 0\\
0& 0& 1 &0& 0& {T_s}\\
0& 0& 0& 1& 0& 0\\
0& 0& 0 &0& 1& 0\\
0& 0& 0& 0& 0& 1
\end{array} \right]
\end{aligned} 
\end{equation}
${{T}_{s}}$ is the sampling interval. Each scatterer is approximated by $C$ particles, so that for the ${{N}_{c}}\left( {{t}_{k}} \right){{M}_{n}}\left( {{t}_{k}} \right)$ scatterers of channel model, a total of ${{N}_{c}}\left( t \right){{M}_{n}}\left( t \right)C$ particles are used for approximation. After that, resampling is needed to extract the particle with the largest weight among the $C$ particles for each scatterer. The weight can be calculated as the Euclidean distance between the $c$th particle state vector and the observation vector $\mathbf{z}\left( {{t}_{k}} \right)$, and it is denoted as
 \begin{equation}
\begin{aligned}
w^{\left( c \right)}\left( {{t}_{k}} \right)=p\left( \left. {{\mathbf{z}}}\left( {{t}_{k}} \right) \right|\mathbf{s}^{\left( c \right)}\left( {{t}_{k}} \right) \right)=\left| {{\mathbf{z}}}\left( {{t}_{k}} \right)-\mathbf{s}^{\left( c \right)}\left( {{t}_{k}} \right) \right|.
\end{aligned} 
\end{equation}
The observation vector is denoted as ${\bf{z}}\left( {{t_k}} \right) = \left[ {{\bf{\tau }}\left( {{t_k}} \right),{{\bf{\alpha }}^T}\left( {{t_k}} \right),{{\bf{\beta }}^T}\left( {{t_k}} \right),{{\bf{\alpha }}^R}\left( {{t_k}} \right),{{\bf{\beta }}^R}\left( {{t_k}} \right)} \right]$, which contains the time-varying channel parameters of the communication channel, including delay, AAoD, EAoD, AAoA, and EAoA.

\section{Statistical properties}
\subsection{Delay Spread}
The delay spread is caused by the time differences of multipath arrival at Rx. 
It is an intuitive and important statistical property in describing the multipath time dispersion of the channel. 
The delay spread is expressed as
 \begin{equation}
\begin{aligned}
{\sigma _\tau }\left( t \right) = \sqrt {{{\bar \tau }^2}\left( t \right) - {{\left( {{\mu _\tau }\left( t \right)} \right)}^2}} 
\end{aligned} 
\end{equation}
where
 \begin{equation}
\begin{aligned}
{\bar \tau ^2}\left( t \right) = \frac{{\sum\limits_{n = 1}^{{N_c}\left( t \right)} {\sum\limits_{m = 1}^{{M_n}\left( t \right)} {{P_{qp,{m_n}}}\left( t \right){{\left( {\tau _{qp,{m_n}}^{NLoS}\left( t \right)} \right)}^2}} } }}{{\sum\limits_{n = 1}^{{N_c}\left( t \right)} {\sum\limits_{m = 1}^{{M_n}\left( t \right)} {{P_{qp,{m_n}}}\left( t \right)} } }} \label{delay_miu}
\end{aligned} 
\end{equation}
 \begin{equation}
\begin{aligned}
{\mu _\tau }\left( t \right) = \frac{{\sum\limits_{n = 1}^{{N_c}\left( t \right)} {\sum\limits_{m = 1}^{{M_n}\left( t \right)} {{P_{qp,{m_n}}}\left( t \right)\tau _{qp,{m_n}}^{NLoS}\left( t \right)} } }}{{\sum\limits_{n = 1}^{{N_c}\left( t \right)} {\sum\limits_{m = 1}^{{M_n}\left( t \right)} {{P_{qp,{m_n}}}\left( t \right)} } }} .\label{delay_sigma}
\end{aligned} 
\end{equation}

\subsection{Angular Spread}
The angular spread describes the degree of dispersion of multipath in space domain. The AAoD (AAoA) and EAoD (EAoA) spreads are defined as
 \begin{equation}
\begin{aligned}
{\sigma _{{\alpha ^{T\left( R \right)}}}}\left( t \right) = \sqrt {{{\bar \alpha }^{T\left( R \right)}}{{\left( t \right)}^2} - {{\left( {{\mu _{{\alpha ^{T\left( R \right)}}}}\left( t \right)} \right)}^2}} 
\end{aligned} 
\end{equation}
 \begin{equation}
\begin{aligned}
{\sigma _{{\beta ^{T\left( R \right)}}}}\left( t \right) = \sqrt {{{\bar \beta }^{T\left( R \right)}}{{\left( t \right)}^2} - {{\left( {{\mu _{{\beta ^{T\left( R \right)}}}}\left( t \right)} \right)}^2}}.
\end{aligned} 
\end{equation}
The calculation of ${{\bar \alpha }^{T\left( R \right)}}$ (${{\bar \beta }^{T\left( R \right)}}$) and ${{\mu _{{\alpha ^{T\left( R \right)}}}}\left( t \right)}$ (${{\mu _{{\beta ^{T\left( R \right)}}}}\left( t \right)}$) is similar to (\ref{delay_miu}) and (\ref{delay_sigma}), replacing ${\tau _{qp,{m_n}}^{NLoS}\left( t \right)}$ with $\alpha _{{m_n}}^{T\left( R \right)}\left( t \right)$ and $\beta _{{m_n}}^{T\left( R \right)}\left( t \right)$, respectively.

\section{Results and analysis}
\begin{figure}[tb]
	\centering{\includegraphics[width=0.35\textwidth]{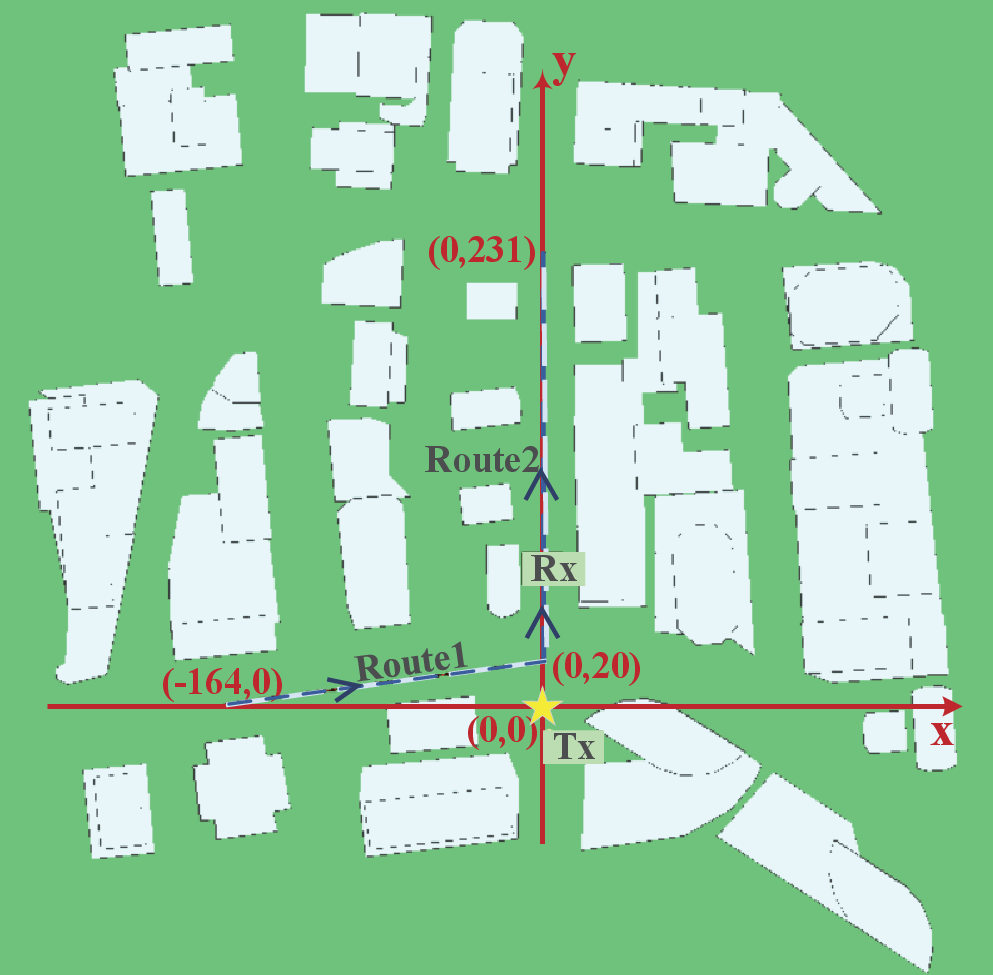}}
	\caption{RT simulation environment map at outdoor scenario.}
	\label{map}
\end{figure}

\begin{figure}[tb]
	\centering
	\subfigure[]{
		\centering
		\includegraphics[width=0.45\textwidth]{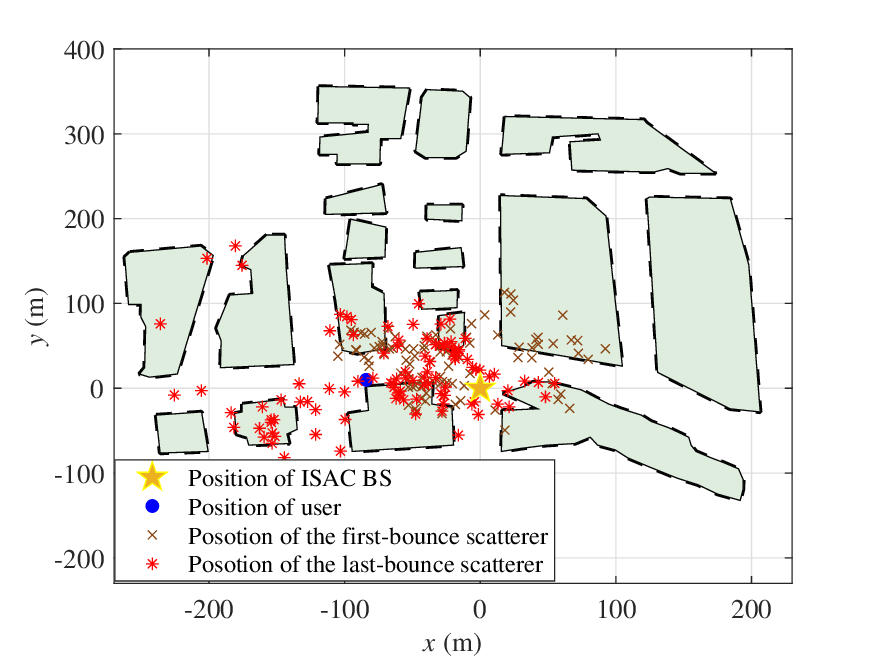}}
	\subfigure[]{
		\centering
		\includegraphics[width=0.45\textwidth]{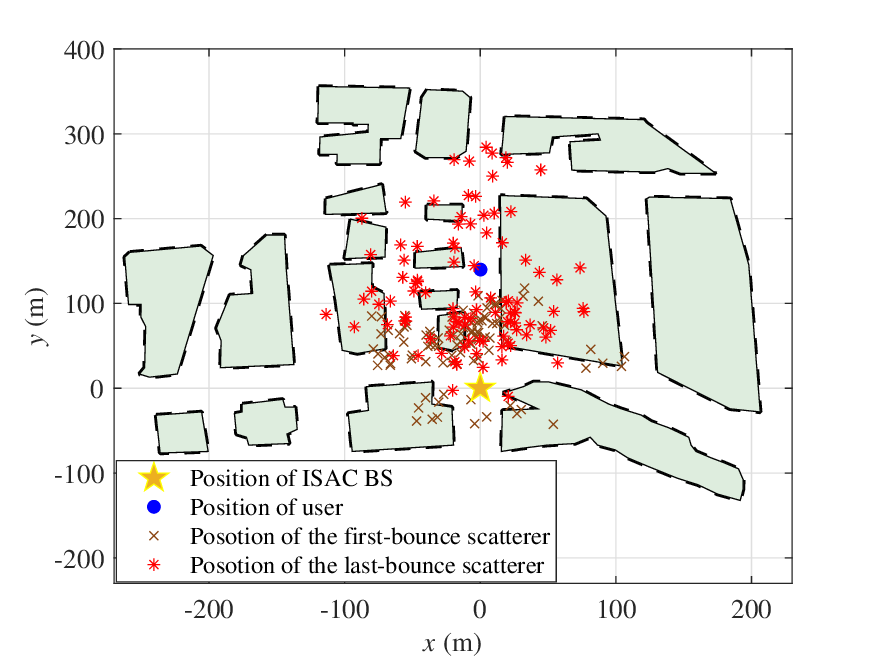}}
	\caption{Estimated locations of user, the first- and last-scatterers 20 s after the initial state, (a) route1 and (b) route2.}
	\label{PlotCluster}
\end{figure}

In this section, the statistical properties of the proposed localization-assisted ISAC channel model are simulated and analyzed.
Because we are not able to conduct the corresponding channel measurements, RT is used to verify the proposed channel model.
The RT simulation is conducted in outdoor scenario at 28 GHz. The Tx equips a $32 \times 4$ planar array antenna, and the Rx equips a $2 \times 2$ planar array antenna. The map of the RT simulation environment is shown in Fig. \ref{map}. Tx is located at the coordinate origin and its height is 10 m. Rx moves at a speed of 1 m/s from route1 to route2. 
The coordinates of the starting point and the ending point at route1 are (-164, 0) and (0, 20), respectively. In addition, route2 is from the starting point (0, 20) to the ending point (0,~231). 
The height of Rx is 1.2 m. 
The simulation results of RT are taken as the observation state of particle filter. When Rx is located at (-104, 7.32, 1.2) of route1 and (0, 120, 1.2) of route2, it is selected as the initial state of the particle filter. The estimated locations of user, the first- and last-scatterers 20~s after the initial state are shown in Fig. \ref{PlotCluster}. It can be found that the scatterers are located around the buildings in the physical environment.
The channel state, including delay and angular parameters, within 20~s after the initial time is estimated by the particle filter. The corresponding statistical properties are compared with the RT simulation results.

\subsection{Delay Spread}
\begin{figure}[tb]
	\centering
	\subfigure[]{
		\centering
		\includegraphics[width=0.23\textwidth]{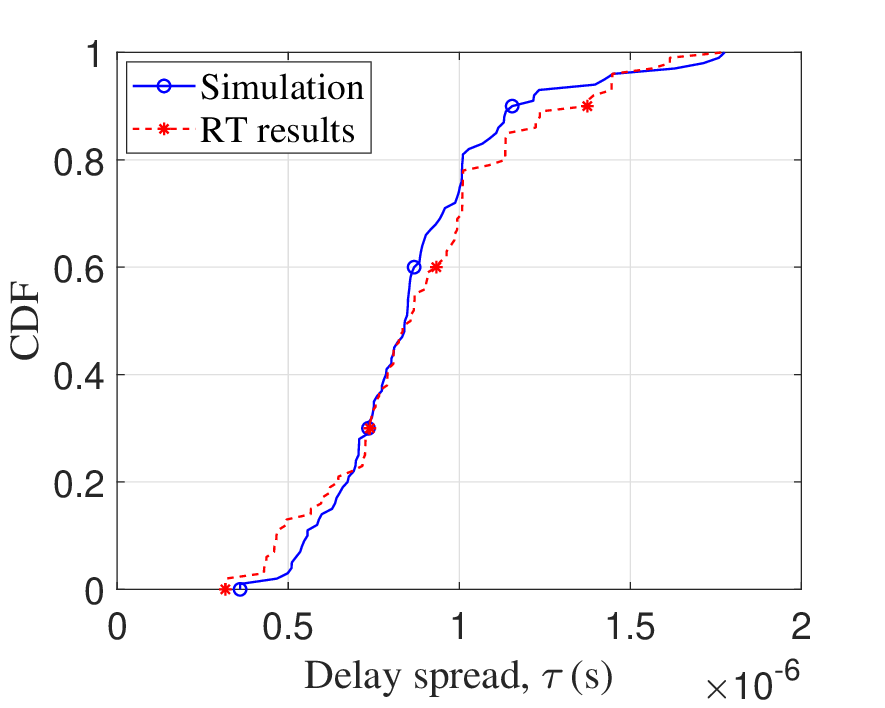}}
	\subfigure[]{
		\centering
		\includegraphics[width=0.23\textwidth]{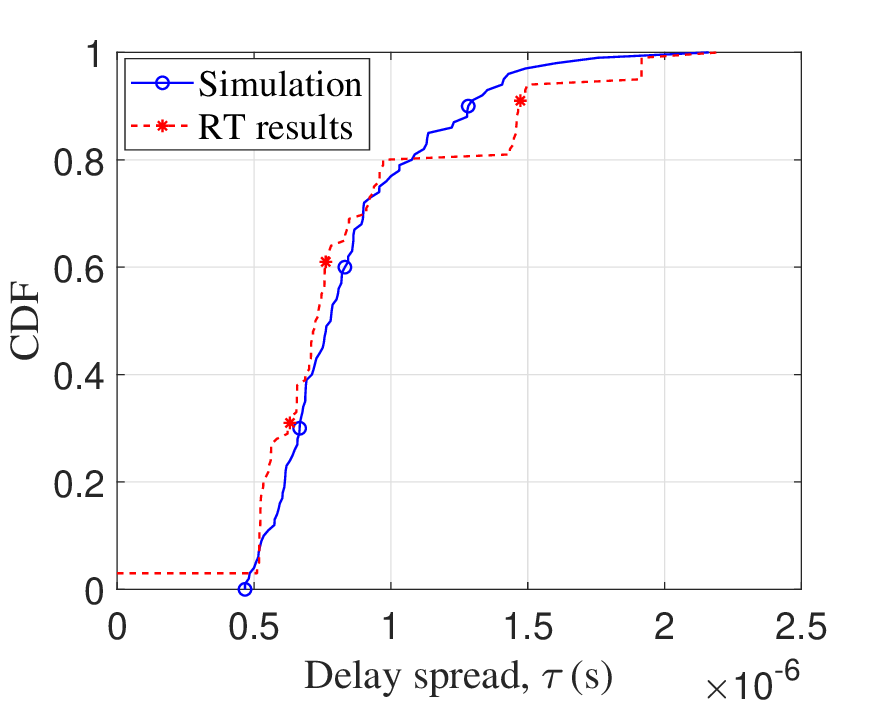}}
	\caption{Cumulative distribution functions (CDFs) of delay spread comparison between ISAC channel model simulation and RT results, (a) route1 and (b) route2.}
	\label{RMSDS_fig}
\end{figure}

Fig.~\ref{RMSDS_fig} compares the delay spreads of localization-assisted ISAC channel model simulation results and RT results.
The delay spreads when Rx at (-84.2, 9.76, 1.2) of route1 and (0, 140, 1.2) of route2 are shown in Fig. \ref{RMSDS_fig} (a) and (b), respectively.
These two points are 20 s after the initial state points.
It shows that the delay spread simulation results of the proposed channel model are almost identical to the RT simulation results for both of these two points.
This indicates that the proposed channel model can estimate and track the multipath delay parameter of the channel.

\subsection{Angular Spread}
\subsubsection{Azimuth Angle Spread}
\begin{figure}[tb]
	\centering
	\subfigure[]{
		\centering
		\includegraphics[width=0.23\textwidth]{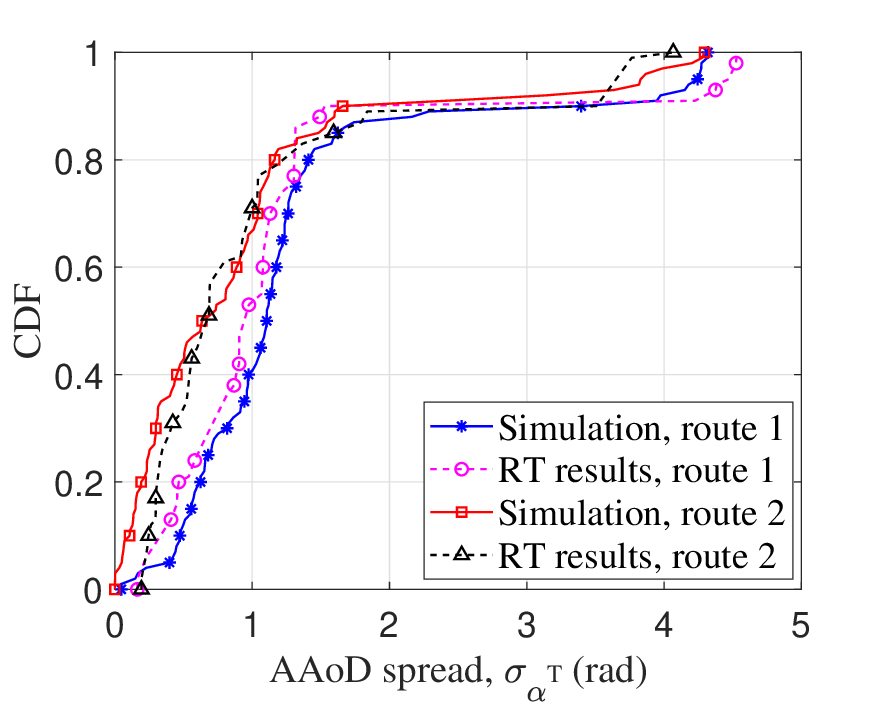}}
	\subfigure[]{
		\centering
		\includegraphics[width=0.23\textwidth]{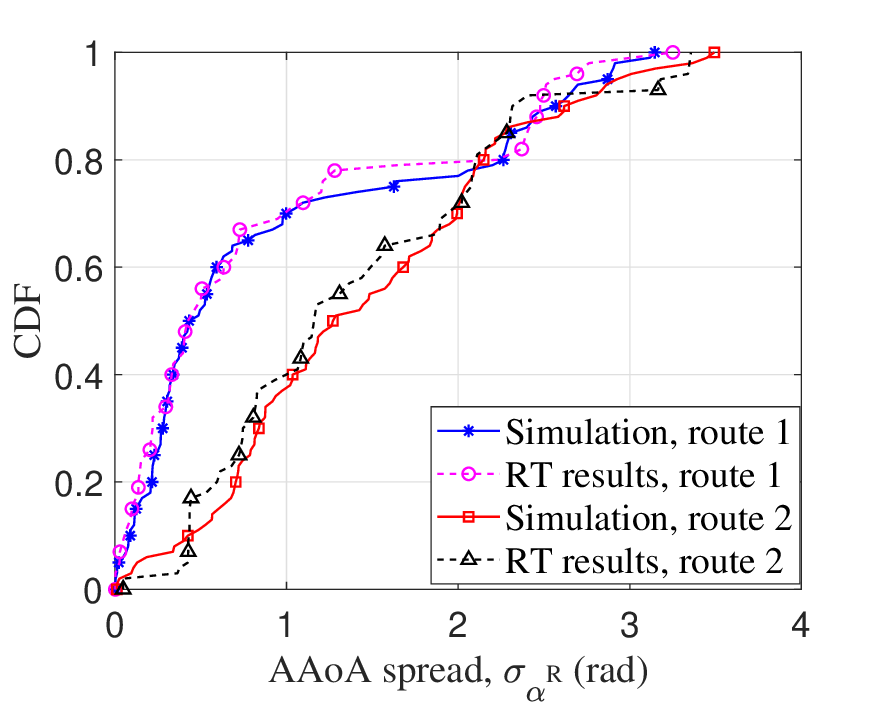}}
	\caption{CDFs of azimuth angle spread comparison between ISAC channel model simulation and RT results, (a) AAoD spread and (b) AAoA spread.}
	\label{AAS_fig}
\end{figure}

Fig. \ref{AAS_fig} (a) and (b) compare the AAoD spread and AAoA spread simulation results of the proposed channel model with that of RT results, respectively. It is found that the azimuth angular spreads of the two points in route1 and route2 both provide a good approximation to the RT results. It shows that the proposed channel model can track and simulate the azimuth parameters well. 

\subsubsection{Elevation Angle Spread}
\begin{figure}[tb]
	\centering
	\subfigure[]{
		\centering
		\includegraphics[width=0.23\textwidth]{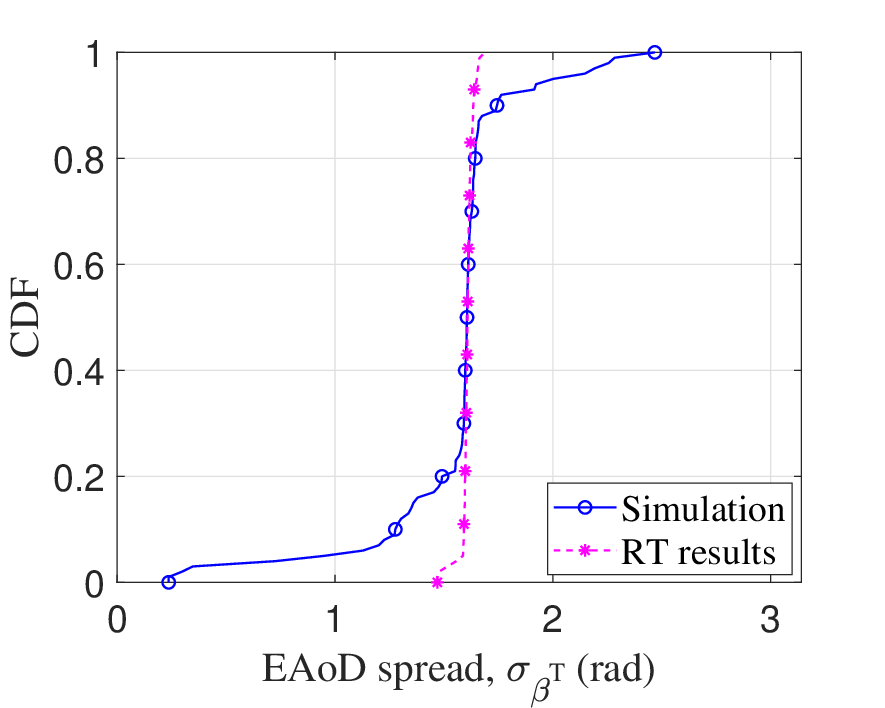}}
	\subfigure[]{
		\centering
		\includegraphics[width=0.23\textwidth]{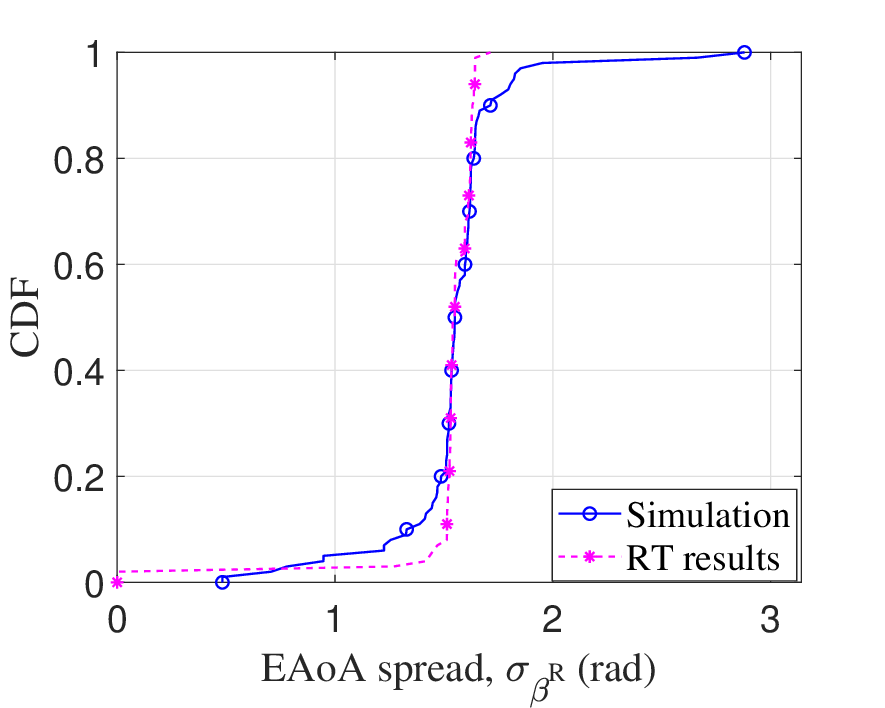}}
	\caption{CDFs of elevation angle spread comparison between ISAC channel model simulation and RT results, (a) route1 and (b) route2.}
	\label{EAS_fig}
\end{figure}

Fig. \ref{EAS_fig} (a) compares the EAoD spread of the point at route1 with the RT results. Fig. \ref{EAS_fig} (b) compares the EAoA spread of the point at route2 with the RT results. The EAoD and EAoA spreads can both match well with the RT results. However, it can be seen that the proposed channel model simulation results have a long tail at both ends of the curve. It may be caused by fewer antenna elements in the elevation direction and insufficient angular resolution.

\section{Conclusions}
In this paper, a novel 3D non-stationary localization-assisted ISAC channel model has been proposed. The particle filter algorithm has been utilized to localize and track the first- and last-bounce scatterers in the channel model. The locations of the scatterers have been used to further assist the communication channel modeling. 
The locations of scatterers estimated by particle filter have shown a difference at different time instances, which explains the non-stationarity of the channel model.
The channel properties have been simulated and compared with RT results, including delay spread, AAoD (AAoA) spread, and EAoD (EAoA) spread. The simulation results of the channel model can accurately fit with RT results, which verifies the correctness of the proposed channel model.

\section*{Acknowledgment}
This work was supported by the National Key R$\&$D Program of China under Grant 2018YFB1801101, the National Natural Science Foundation of China (NSFC) under Grants 61960206006 and 62271147, the Key Technologies R$\&$D Program of Jiangsu (Prospective and Key Technologies for Industry) under Grants BE2022067 and BE2022067-1, the Frontiers Science Center for Mobile Information Communication and Security, the EU H2020 RISE TESTBED2 project under Grant 872172, and the High Level Innovation and Entrepreneurial Doctor Introduction Program in Jiangsu under Grant JSSCBS20210082.

\bibliographystyle{IEEEtran}

\begin{thebibliography}{00}
	\bibitem{b1} X.-H. You, C.-X. Wang, J. Huang, \emph{et al.}, ``Towards 6G wireless communication networks: Vision, enabling technologies, and new paradigm shifts,” \emph{Sci. China Inf. Sci.}, vol. 64, no. 1, pp. 1--74, Jan. 2020, doi: 10.1007/s11432-020-2955-6.
	 
	\bibitem{rrb1} C.-X. Wang, J. Wang, S. Hu, \emph{et al.}, ``Key technologies in 6G THz wireless communication systems: A survey,” \emph{IEEE Veh. Technol. Mag.}, vol. 16, no. 4, pp. 27–-37, Dec. 2021. 
	
	\bibitem{b2} Z. Feng, Z. Fang, Z. Wei, X. Chen, Z. Quan, and D. Ji, ``Joint radar and communication: A survey," \emph{China Commun.}, vol.~17, no.~1, pp.~1--27, Jan.~2020.
	\bibitem{rb1} J. A. Zhang, Md. Lushanur Rahman, K. Wu, \emph{et al.}, ``Enabling joint communication and radar sensing in mobile networks—A survey," \textit{IEEE Commun. Surveys Tuts.}, vol.~24, no.~1, pp.~306--345, 1th Quart.~2022.
	
	\bibitem{rb2} F. Liu, C. Masouros, A. P. Petropulu, H. Griffiths, and L. Hanzo, ``Joint radar and communication design: Applications, state-of-the-art, and the road ahead," \textit{IEEE Trans. Commun.}, vol.~68, no.~6, pp.~3834--3862, June~2020.
	

	
	\bibitem{b3} C.-X. Wang, Z. Lv, X. Gao, \emph{et al.}, ``Pervasive wireless channel modeling theory and applications to 6G GBSMs for all frequency bands and all scenarios,”  \textit{IEEE Trans. Veh. Technol.}, vol. 71, no. 9, pp. 9159--9173, Sept. 2022. 
	
	\bibitem{b5} H. Deng and A. Sayeed,``Mm-wave MIMO channel modeling and user localization using sparse beamspace signatures," in \textit{Proc. SPAWC’14}, Toronto, ON, Canada, June~2014, pp.~130--134.
	\bibitem{b6} J. Rodríguez-Piñeiro, Z. Huang, X. Cai, T. Domínguez-Bolaño, and X. Yin, ``Geometry-based MPC tracking and modeling algorithm for time-varying UAV channels," \textit{IEEE Trans. Wireless Commun.}, vol.~20, no.~4, pp.~2700--2715, Apr.~2021.
	\bibitem{b7} M. Zhu, J. Vieira, Y. Kuang, K. Åström, A. F. Molisch, and F. Tufvesson,~``Tracking and positioning using phase information from estimated multi-path components," in \textit{Proc. ICCW'15}, London, UK, June~2015, pp.~712--717.
	
	\bibitem{b8} Y. Ji, J. Hejselbæk, W. Fan, and G. F. Pedersen, ``A map-free indoor localization method using ultrawideband large-scale array~systems," \textit{IEEE Antennas Wireless Propag. Lett.}, vol. 17, no. 9,~pp.~1682--1686, Sept.~2018.
	
	\bibitem{b9} A. Guerra, F. Guidi, D. Dardari, A. Clemente, and R. D’Errico, ``A millimeter-wave indoor backscattering channel model for environment mapping," \textit{IEEE Trans. Antennas Propag.}, vol.~65, no.~9, pp.~4935--4940, Sept.~2017. 
	
		
	\bibitem{b11} Q. Zhu, K. Jiang, X. Chen, W. Zhong, and Y. Yang, ``A novel 3D non-stationary UAV-MIMO channel model and its statistical properties,” \textit{China Commun.}, vol. 15, no. 12, pp. 147--158, Dec. 2018.
	
	\bibitem{b12} R. He, B. Ai, G. L. Stüber, and Z. Zhong, ``Mobility model-based non-stationary mobile-to-mobile channel modeling,” \textit{IEEE Trans. Wireless Commun.}, vol. 17, no. 7, pp. 4388--4400, July 2018.
	 
	
	
	
	
	\bibitem{c1} C. Ge, R. Zhang, Y. Jiang, B. Li, and Y. He, ``A 3-D dynamic non-WSS cluster geometrical-based stochastic model for UAV MIMO channels," \textit{IEEE Trans. Veh. Technol.}, vol. 71, no. 7, pp. 6884--6899, July 2022.
	
	\bibitem{c2} J. Bian, C.-X. Wang, Y. Liu, \emph{et al.}, ``3D non-stationary wideband UAV-to-ground MIMO channel models based on aeronautic random mobility model,” \textit{IEEE Trans. Veh. Technol.}, vol. 70, no. 11, pp. 11154--11168, Nov. 2021. 
	
	\bibitem{c3} Y. Bi, J. Zhang, Q. Zhu, \emph{et al.}, ``A novel non-stationary high-speed train (HST) channel modeling and simulation method,” \textit{IEEE Trans. Veh. Technol.}, vol. 68, no. 1, pp. 82--92, Jan. 2019.
\end{thebibliography}

\end{document}